\documentclass[10pt]{article}
\usepackage{amsmath}
\usepackage{amssymb}
\usepackage{graphicx}
\usepackage{cite}
\usepackage{color} 
\usepackage[labelfont=bf,labelsep=period,justification=raggedright]{caption}
\makeatletter
\renewcommand{\@biblabel}[1]{\quad#1.}
\makeatother

\date{}

\begin{document}

\begin{flushleft}
{\Large
\textbf{Dynamics of p53 and Wnt cross talk.}
}
\bigskip
\\
Md.Zubbair Malik$^{1,3}$, Shahnawaz Ali$^{1,3}$,Md. Jahoor Alam$^{2,3}$,Romana Ishrat$^1$, R.K. Brojen Singh$^{*3}$
\\
\bigskip
$^1$Centre for Interdisciplinary Research in Basic Sciences, Jamia Millia Islamia, New Delhi-110025, India.
\\
$^2$College of Applied Medical Sciences, University of Ha’il, P.O. Box 2440, Ha’il, Kingdom of Saudi Arabia.
\\
$^3$School of Computational and Integrative Sciences, Jawaharlal Nehru University, New Delhi-110067, India
\\
\bigskip
$\ast$ Corresponding author, E-mail:  R.K. Brojen Singh - brojen@jnu.ac.in, \\

\end{flushleft}

\section*{Abstract}
We present the mechanism of interaction of Wnt network module, which is responsible for periodic sometogenesis, with $p53$ regulatory network, which is one of the main regulators of various cellular functions, and switching of various oscillating states by investigating $p53-$Wnt model. The variation in Nutlin concentration in $p53$ regulating network drives the Wnt network module to different states, stabilized, damped and sustain oscillation states, and even to cycle arrest. Similarly, the change in Axin concentration in Wnt could able to modulate the $p53$ dynamics at these states. We then solve the set of coupled ordinary differential equations of the model using quasi steady state approximation. We, further, demonstrate the change of $p53$ and $Gsk3$ interaction rate, due to hypothetcal catalytic reaction or external stimuli, can able to regulate the dynamics of the two network modules, and even can control their dynamics to protect the system from cycle arrest (apoptosis).

\bigskip
{\bf Keywords:} $p53$ activation, Fixed point oscillations, Nutlin,Wnt, Sustain oscillations.

\section*{Introduction}
$p53$, one of the largest hub in cellular network \cite{col,mos}, is considered to be one of the most important key regulators of cellular metabolic pathways \cite{Ita,Vog}, regulates a number of cellular functions \cite{aga} and its dynamics control even the fate of the cell when stress is given to the cell \cite{pur}. Further, $p53$ suppression is observed in various types of cancer either due to mutations or by ambiguous expression of control systems like $MDM2$ (or $HDM2$ its human equivalent) \cite{Low}. It is negatively regulated by the $MDM2$ \cite{moma}, and functions as an $E3$ ligase which mediates the proteosomic degradation of $p53$ \cite{Tol,Hon}, facilitates the diffusion of $p53$ in the nucleus \cite{Kob}, which in turn lowers the level of $p53$ within the cell \cite{Oli}. In stressed cells, the $p53$ triggers the cell to cell cycle arrest forcing it to choose its fate, either to repair or apoptosis \cite{Che,Hau,aro}.  

$p53$ dynamics is regulated by various signaling molecules as evident from various experimental and theoretical reports \cite{ore}. On of the most important inhibitor of $p53-MDM2$ is nutilin \cite{san}. It is a selective small molecule which directly binds to binding pocket of $MDM2$ \cite{Mom}, to activate $p53$ pathway \cite{Mom,Vas}, and inhibiting $p53$ binding to $MDM2$ \cite{Log}. $Nutilin-3a$ and $3b$ are the two most active enantiomorphs that are found to up-regulate the $p53$ in $p53$ dependent manner \cite{Vil}. Further, it is also demonstrated that $nutlin$-$3$ induces anti-angiogenic activities in endothelial cells probably via three mechanisms; first by inhibiting endothelial cell migration; second by inducing cell cycle arrest; and third by increasing apoptotic tendency in endothelial cells \cite{bin}. Further, it was also shown that $nutlin$-$3$ treatment in these cells leads to accumulation of $p53$ \cite{Lee}, indicating the important impacts of $nutlin$-$3$ which interferences the $p53$-$MDM2$ regulatory mechanism. As a consequence, it was shown that $Nutilin$-$3$ stabilizes the $p53$ dynamics and causes the activation of survival pathways \cite{Lee}.

Somitogenesis in vertebrates is periodic formation of somites (vertabrae precursors) in the anterior presomitic mesoderm tissue (PSM) controlled by complex gene network known as segmentation clock \cite{aul}, where, Notch, Wnt pathways \cite{gib1,gib2} and fibroblast growth factor (FGF) are the main components \cite{Gol}. Further, it has been shown that these three main pathways generate rhythms of specific oscillation relationships \cite{Gol}, and cross-talk among them \cite{deq} as a basis of spatio-temporal self-organization of patterns during sometogenesis \cite{pou}. This physiological oscillations is responsible for periodic spacing of somites, dynamic structures and regular segmented development in vertebrates \cite{pou}. In some studies in normal (stress) and cancerous cells, it has been reported that $p53$ and Wnt modules are being coupled and cross-talk between them via different intermediate proteins or genes or signaling molecules \cite{pen,Cha,Kim}. The studies also reported that $p53$ network suppresses the Wnt signaling cascade \cite{Kim}. The Wnt signaling plays a pivotal role in determination of cell fate, which may probably through $p53$ interaction \cite{Kim,Cha}, but still it is not clear yet. Further, several other works reported that the intercellular signaling of Wnt depends on the $\beta$-catenin and glycogen synthase $kinase$-$3$ ($GSK$-$3$) \cite{Tau}. Moreover, $p53$ can regulate the dynamics of $Axin2$ and $GSK$-$3$ \cite{Kimn}. Recently, it is also reported that the $miR-34$ family, which is directly transcribed by the $p53$, links $p53$ with Wnt signaling \cite{Cha}, in this process, and a set of $GSK$-$3$-related Wnt genes, are directly targeted by $p53$ and $miR$-$34$. These observations clearly suggest the close connection between the $p53$ functionalities and Wnt in stress and cancer cells \cite{pen,Cha}. However, the way how $p53$ interacts with Wnt during sometogenesis and its impact on somite organization/reogranization at molecular level is still an open question.

$GSK$-$3$, which is an important coupling molecule between $p53$ and Wnt pathways, is widely expressed in broad range of cellular processes and being a Ser/Thr kinase it is a multifunctional protein \cite{Dob,Woo}. It is active under resting condition and act as a core regulator in various disease pathways like cancer \cite{Dob}. Being a kinase, it confers selectivity and substrate specificity \cite{Dob}. It also largely acts as a phosphorylating and an important signaling agent for the degradation of large number of proteins, such as $\beta$-catenin in Wnt pathway \cite{Abe}. Further, it is a central player in Wnt signaling pathway recruited with the $Axin2$ at the Frizzle receptor for the activation of Wnt signaling \cite{Nak}. In absence of Wnt signals $GSK$-$3$ shows its activity in the protein destruction complex with $Axin2$, APC (adenomatous polyposis coli) and other partners mediating the destruction and phosphorylation of $\beta$-catenin \cite{Tae}. $GSK$-$3$ also mediates $p53$ dependent apoptosis signaling by suppressing $p53$-AKT pathway. The role $GSK-3$ as a mediator of $p53$ and Wnt cross-talk is not fully understood. The impact of other Wnt regulators in Wnt pathway and how they regulate $p53$ dynamics, and vice versa, are needed to be investigated in order to understand how segmentation clock works.

We study the coupling of Wnt oscillator which is one of the three important clocks in sometogenesis in vertabrates and $p53$ regulatory network to understand how does these two oscillators regulates each other and process information during somite formation and pattern organization. Since $p53$ interferes various other cellular functions, it will be interesting to understand its regulating impacts on segmentation clock and vice versa. Presently we study these phenomena by modeling this $p53$-Wnt two oscillator model with updated interaction between them from various experimental reports and how do they regulate each other. We organize our work as follows. The detailed description of the model and techniques is provided in materials and methods section. It is followed by mathematical treatment of the model, the numerical simulation results and its discussion in the next section. We then draw come conclusions based on the results we obtained and discussion on it.

\section*{Materials and Methods}

\subsection*{A. Coupling p53-Wnt Oscillators Model}
The Wnt signaling pathway is authorized by $\beta$-catenin regulation via various intermediate interaction steps (Fig. 1) \cite{log1,Jen}. In this signal processing, Wnt promotes the synthesis of Dishevelled (Dsh) protein, and then inhibits $GSK3$ protein kinase \cite{Gol}. $GSK3$, then interacts with $Axin$ (with rates $k_1$ and $k_2$ for forward and backward reactions as in Fig.1) in the presence of $\beta$-catenin to form transient complex which in turn phosphorylates $\beta$-catenin into $\beta$-catenin-P (with rate rate $k_6$) \cite{Gol,Jen}. The $\beta$-catenin-P degraded quickly with rate $k_{14}$, on the other hand it is dephosphorylated with rate $k_3$ to form freed $\beta$-catenin and then degraded with a rate $k_{13}$ \cite{Cle}. The free $\beta$-catenin is then transported to nucleus ($\beta$-catenin-n) with rate $k_4$, on the other hand $\beta$-catenin-n is transpoted from nucleus to cytoplasm with another rate $k_5$. Then $\beta$-catenin-n promotes the transcription of set of target genes including mAxin (mRNA of Axin, specifically homolog Axin2 in PSM tissue) at the rate $k_{15}$, with rate $k_{16}$, and mAxin also degrades with a rate $k_{18}$ \cite{Hue}. On the other hand mAxin translates to Axin with a rate Axin protein which also degrades with a rate $k_{23}$. This mechanism of Axin is just like a negative feedback loop which drives the oscillations in the network components. Thus $Axin2$ is the essential target of the Wnt pathway because it inhibits the Wnt signals by degrading the $\beta$-catenin by forming a negative feedback loop \cite{Wan}. 
\begin{figure*}
\label{fig1}
\begin{center}
\includegraphics[height=350pt,width=13.0cm,angle=270]{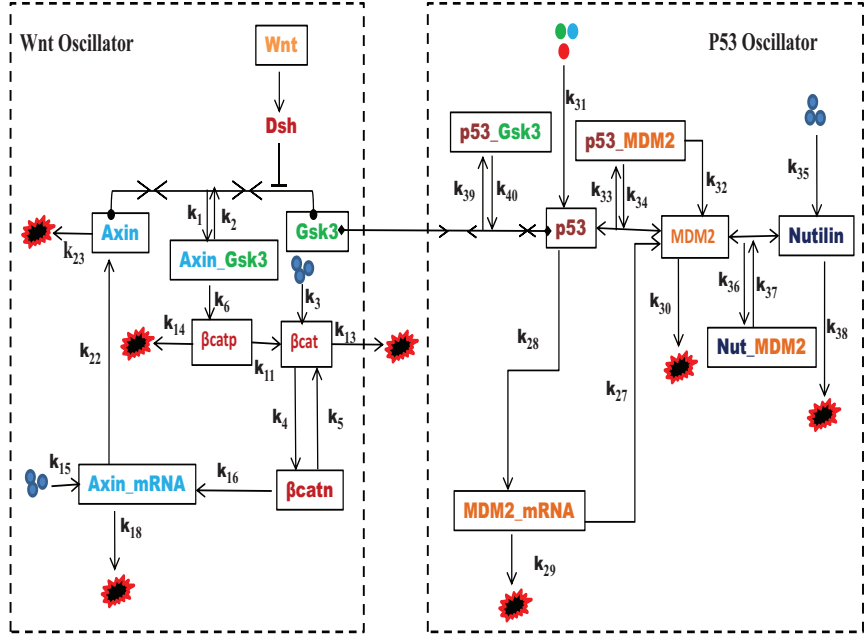}
\caption{The schematic diagram of biochemical network of $p53$ and Wnt modules and their cross talk via GSK3.} 
\end{center}
\end{figure*}

The Wnt pathway communicates with the $p53$ Oscillator \cite{Pro,Ala,Dev} through $GSK3$ by formation of a binary complex $p53-GSK3$ with the free available $p53$ at the rate $k_{39}$. Then this complex $p53-GSK3$ degrades at the rate $k_{40}$ to the free $p53$ and $GSK3$ proteins \cite{Proc}. $p53$ and $MDM2$ regulate each other with feedback mechanism where formation and decay of various intermediate complexes are involved with different rates (Fig. 1) as described in Table 1 and 2 \cite{Pro}. The freely synthesized Nutilin directly interacts with $MDM2$ with a rate $k_{36}$ to form a complex $Nut_MDM2$, and then this complex dissociates to release back free Nutlin and $MDM2$. Then Nutilin indirectly interacts with $p53$ via $MDM2$ by affecting the feedback mechanism between $p53$ and $MDM2$. The Nutilin then degrades with rate constant $k_{38}$. Thus the activated $p53$ can now go to the nucleus and initiate the transcription of a number of target genes including $MDM2$. $MDM2$ is essential in controlling the concentration of $p53$ through a positive feedback loop mechanism by making complex \cite{Chi,Lah}. 

\subsection*{Mathematical model of the network: Quasi steady state approximation}
The stress $p53-Wnt$ regulating network we study is defined by $N=13$ molecular species (Table 1) corresponding to the reaction network description provided in Table 2. The state of the system at any instant of time $t$ is given by the state vector, $\vec{x}(t)=\left(x_1,x_2,\dots,x_N\right)^T$, where, $N=13$ and $T$ is the transpose of the vector. By considering feedback mechanism of in $p53$ and Wnt oscillators and coupling reaction channels of the two oscillators, we could able to reach the following coupled ordinary differential equations (ODE),
\begin{eqnarray}
\label{ode}
\frac{d}{dt}\vec{x}(t)=\left[\begin{matrix}F_1\\F_2\\ \vdots\\F_{13}\end{matrix}\right]
\end{eqnarray}
where, the functions in the equation (\ref{ode}) $\{F_i(x_1,x_2,\dots,x_N)\},i=1,2,\dots,13$ are given by,
\begin{eqnarray}
\label{deter}
F_1&=& k_{26}\left[k_{22}x_2-k_{23}\frac{x_1}{k_{24}+x_1}-k_1x_1x_6+k_2\left(k_8-x_6\right)\right] \ \\
F_2&=& k_{26}\left[k_{15}+k_{16}\frac{\left(x_5\right)^{k_{25}}}{\left(k_{17}\right)^{k_{25}}+\left(x_5\right)^{k_{25}}}-k_{18}\left(\frac{x_2}{k_{19}+x_2}\right)\right] \\
F_3&=& k_{26}\left[k_3-k_6\left(\frac{k_9}{k_9+k_7}\right)\left(\frac{x_3}{k_{10}+x_3}\right)\left(\frac{k_8-x_6}{k_8}\right)+k_{11}\left(\frac{x_4}{k_{12}+x_4}\right)\right]\nonumber\\
&&-k_{26}\left[k_4x_3-k_5x_5+k_{13}x_3\right]\\
F_4&=& k_{26}\left[k_6\left(\frac{k_9}{k_9+k_7}\right)\left(\frac{x_3}{k_{10}+x_3}\right)\left(\frac{k_8-x_6}{k_8}\right)-k_{11}\left(\frac{x_4}{k_{12}+x_4}\right)-k_{14}x_4\right]\\
F_5&=& k_{26}\left[k_4x_3-k_5x_5\right]
\end{eqnarray}
\begin{eqnarray}
F_6&=& k_{26}\left(-k_1x_1x_6+k_2k_8-k_2x_6\right)-\left(k_{39}x_7x_6+k_{40}x_{13}\right)\\
F_7&=& k_{31}-k_{33}x_7x_8+k_{34}x_{10}-k_{39}x_7x_6+k_{40}x_{13}\\
F_8&=& k_{27}x_9-k_{30}x_8+k_{32}x_{10}-k_{33}x_7x_8+k_{34}x_{10}-k_{36}x_{11}x_8\\
F_9&=& k_{28}x_7-k_{29}x_9+k_{41}x_{13}\\
F_{10}&=& -k_{32}x_{10}+k_{33}x_7x_8-k_{34}x_{10}\\
F_{11}&=& k_{35}-k_{36}x_{11}x_8+k_{37}x_{12}-k_{38}x_{11}\\
F_{12}&=& k_{36}x_{11}x_8-k_{37}x_{12}\\
F_{13}&=& k_{39}x_7x_6-k_{40}x_{13}
\end{eqnarray}
The coupled ODEs (equation (\ref{ode})) are very difficult to solve analytically. However, one can get approximate analytical solution of these equations by using quasi-steady state approximation \cite{Mur,Sch}. In general, any biochemical reactions network involves two basic types of reaction, namely slow and fast reactions \cite{Mur}. Therefore, the $N=13$ variables in the system can be divided into sets of slow and fast variables respectively. If $\vec{x}^s(t)=\left(x_1,x_2,\dots,x_l\right)^T$ and $\vec{x}^f(t)=\left(x_{l+1},x_{l+2},\dots,x_N\right)^T$ are slow and fast variable vectors, then $\vec{x}(t)=\left(\vec{x}^s,\vec{x}^f\right)^T$. Then from equation (\ref{ode})-(14) along with Table I, we have,
\begin{eqnarray}
\label{steady}
\vec{x}^s(t)=\left[\begin{matrix}x_{1}\\x_{3}\\x_{6}\\x_{7}\\x_{8}\\x_{11}\\\end{matrix}\right];~~~
\vec{x}^f(t)=\left[\begin{matrix}x_{2}\\x_4\\x_{5}\\x_{9}\\x_{10}\\x_{12}\\x_{13} \end{matrix}\right]
\end{eqnarray}
Since the rate of complex formation is fast, and after this fast complex formation, the fast variables immediately retain steady state (equilibrium). Then using Henri-Michaelis-Menten-Briggs-Haldane approximation \cite{Ped}, one can take quite fair assumption that the ODEs of variables of complex molecular species reach steady state equilibrium quite fast compared with the time evolution of slow variables \cite{Mur}. Then one can straight forward put, $\frac{d\vec{x}^f}{dt}\approx 0$ \cite{Ped,Mur}. Now, the number of coupled ODE (\ref{ode}) reduces to the following,
\begin{eqnarray}
\label{reduce}
\frac{\vec{x}(t)}{dt}=\frac{\vec{x}^s(t)}{dt}=\frac{d}{dt}\left[\begin{matrix}x_{1}\\x_{3}\\x_{6}\\x_{7}\\x_{8}\\x_{11}\\\end{matrix}\right]
\end{eqnarray}
and the $\vec{x}^f$ become the following steady state values,
\begin{eqnarray}
\label{equi}
\vec{x}^s\rightarrow\left[\begin{matrix}x_{2}^*\\x_4^*\\x_{5}^*\\x_{9}^*\\x_{10}^*\\x_{12}^*\\x_{13}^*\end{matrix}\right]\rightarrow constant
\end{eqnarray}
The thirteen ordinary differential equations are now reduced to six ordinary differential equations. This allows to simplify the complex mathematical model to get approximate solutions of the system numerically saving computational cost or analytically fron the reduced system if possible.

We used standard Runge-Kutta method (order 4) of numerical integration to simulate equation (ref{ode}) to find the solution of the variables listed in Table 1 for the parameter values given in Table 2. We then analysed the constructed mathematical model to get possible approximate analytical solutions of the variables (slow variables) using quasi-steady state approximation.
\begin{figure*}
\label{fig2}
\begin{center}
\includegraphics[height=240pt,width=12.0cm]{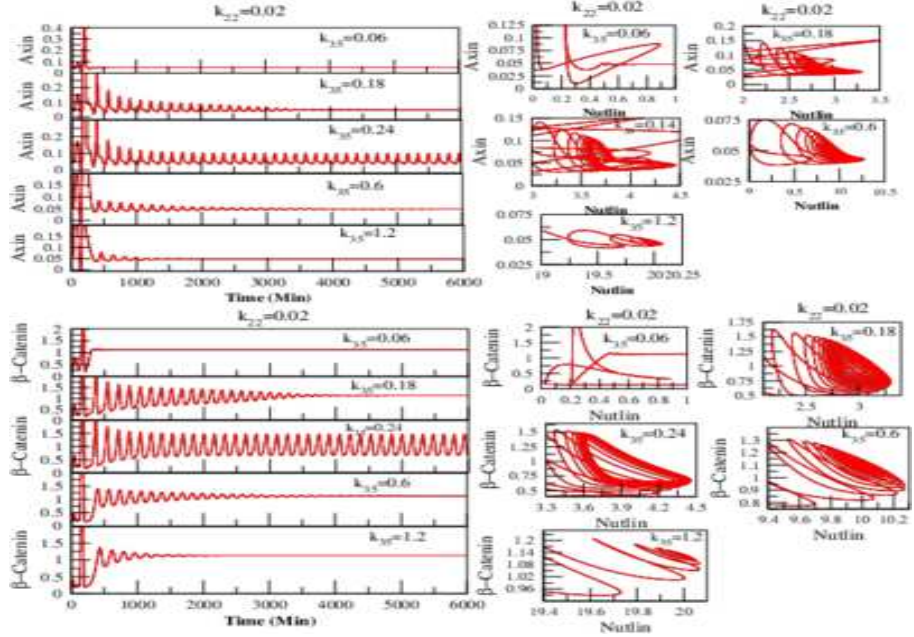}
\caption{The dynamics of Axin and $\beta$-catenin for different values of $k_{35}$ i.e. 0.06, 0.18, 0.24, 0.6 and 1.2 for fixed vale of $k_{22}=0.02$. The right hand panels are the two dimensional plots the Axin and $\beta$-catenin with nutlin for the same set of parameter values showing different state behaviors.} 
\end{center}
\end{figure*}

\section*{Results and Discussion}

The cross talk between $p53$ and Wnt via Gsk3 subjected to different stress conditions induced by nutlin, as well as Axin concentrations available in the system. In order to understand how they regulate each other, we numerically simulated the coupled ordinary differential equations (\ref{ode}) with the parameters listed in Table 2. The simulation results for the variable in the model are presented and compared to understand the switching of different oscillating states induced by different stress conditions. The reason is that these states of the oscillations in Wnt network are responsible for different behavior of regular somite formation, and many other periodic phenomena during sometogenesis. The roles of $Axin2$, $Gsk3$, $\beta$-catenin and Nutilin on the $p53$ network dynamics is discussed in the context of the model we presented. Further, the signal processing between the two coupled network modules via the $Gsk3$ during sometogenesis is studied under the stress $p53$ generating effect due to $Nutilin$-$3a$. On the other hand, we also studied the regulation of $p53$ by Wnt due to available Axin concentration in the system.

\subsection*{Driving Wnt oscillating states by $p53$}
The $p53$ in $p53$ regulatory network can be induced stress by the available concentration of stress inducer molecular species, $Nutlin-3a$. 
The concentration level of $Nutlin-3a$ is proportional to the creation rate constant of $Nutlin-3a$, $k_{35}$ (Fig. 1). Therefore, computationally we can able to monitor the amount of stress induced to $p53$ by changing the value of parameter $k_{35}$ and supervising the dynamical behavior of $p53$ (Fig. 2 and 3). 

The dynamics of $p53$ (Fig. 2 upper left panels) for small values of $k_{35}$ ($k_{35}\langle 0.07$) show single spike due to sudden stress, and due to small $k_{35}$ values, the dynamics become stabilized indicating normal nature of the system. Then, as the value of $k_{35}$ increases, $p53$ dynamics show damped oscillation indicating the inducing oscillation for certain interval of time ($t\approx [0,4000]$ minutes for $k_{35}=0.18$), and then its dynamics become stabilized. The time interval of damped oscillation regime depends on the value of $k_{35}$ and increases as $k_{35}$. This indicates that if the stress given to the system is small, the system first will go to the stress or excited state (indicated by oscillatory behavior), repair back the changes in the system and come back to the normal condition. For sufficiently large values of $k_{35}$ ($0.23\le k_{35}\le 0.55$) the $p53$ dynamics become sustain oscillation, where the responsed stress is optimal and sustains the stress for $t\rightarrow\infty$. Further increase in the value of $k_{35}$ ($k_{35}>0.55$) force the sustain oscillation of $p53$ to damped oscillation whose time interval of damped oscillation decreases as $k_{35}$ increases. This indicates that excess stress in $p53$ due to Nutlin may become toxic to the system reflected in $p53$ dynamics. If we increase the value of $k_{35}$ ($k_{35}>1.3$) the $p53$ dynamics become stabilized. This may probably the case of apoptosis of the system. The results obtained are in consistent with the experimental observations which indicates that acetylation of $p53$ is responsible for its activation and stabilization \cite{Gu}. If we further increase the value of $k_{35}$, $p53$ activation decreases maintaining $p53$ stability, but at higher values ($\le$35).
\begin{figure*}
\label{fig3}
\begin{center}
\includegraphics[height=280pt,width=13.0cm]{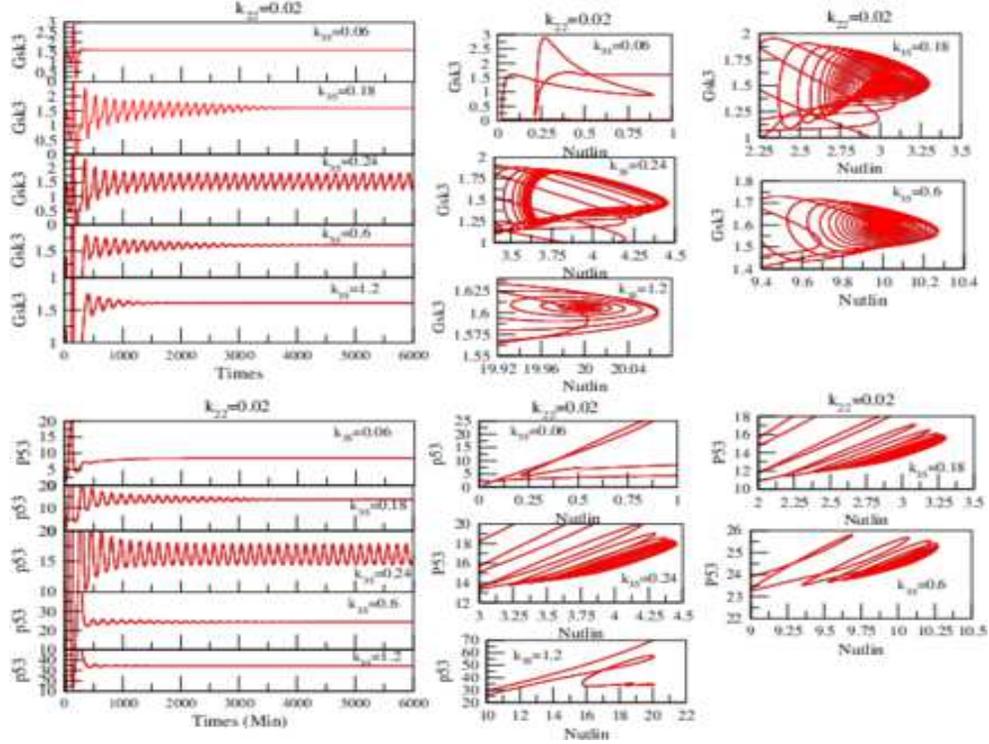}
\caption{The similar plots for $p53$ and GSK3 for the same set of values as in Figure 2.} 
\end{center}
\end{figure*}

The two dimensional plots between $p53$ and Nutlin (Fig. 2 upper right panels) for different values of $k_{35}$ show different behaviors of the system, namely stable fixed point, damped oscillation towards a fixed point, sustain oscillation, then damped and stabilized fixed point. This behaviors show the same behaviors as shown in $p53$ dynamics for different values of $k_{35}$.

Similarly, we study the dynamics of $Gsk3$ as a function of $k_{35}$ (Fig. 2 lower left panels) for corresponding values which we took for the case of $p53$. In this case also we found the similar behaviors as we got in $p53$ case (Fig. 2), namely, stabilized, damped, sustain, then damped and stabilized states for the corresponding values of $k_{35}$ taken in $p53$ case. The two dimensional plots of $Gsk3$ and Nutlin (Fig. 2 lower right hand panels) show the evidence of the existence of the mentioned states.

Now we study the dynamics of Axin (Fig. 3 upper left panel) imparted by stress $p53$ via $Gsk3$ induced by Nutlin concentration in the system. The signal of the stress $p53$ is processed by Axin via $Gsk3$ and reflected the same states in Axin dynamics as obtained in $p53$. The Axin and $\beta-Catenin$ dynamics as a function of $k_{35}$, as well as two dimensional plots of these two molecules with Nutlin are similar qualitatively with the dynamics of stress $p53$ and $Gsk3$ (Fig. 2 and 3). This means that depending on the values of $k_{35}$ the stress $p53$ can able to generate and arrest the Axin and $\beta-Catenin$ cycles. Thus the dynamics of Axin and $\beta-Catenin$ can be regulated by $p53$ and probably can control the Wnt oscillator. This means that $p53$ regulation has strong impact on Wnt oscillation, and regulate various mechanisms during sometogenesis.
\begin{figure*}
\label{fig4}
\begin{center}
\includegraphics[height=250pt,width=13.0cm]{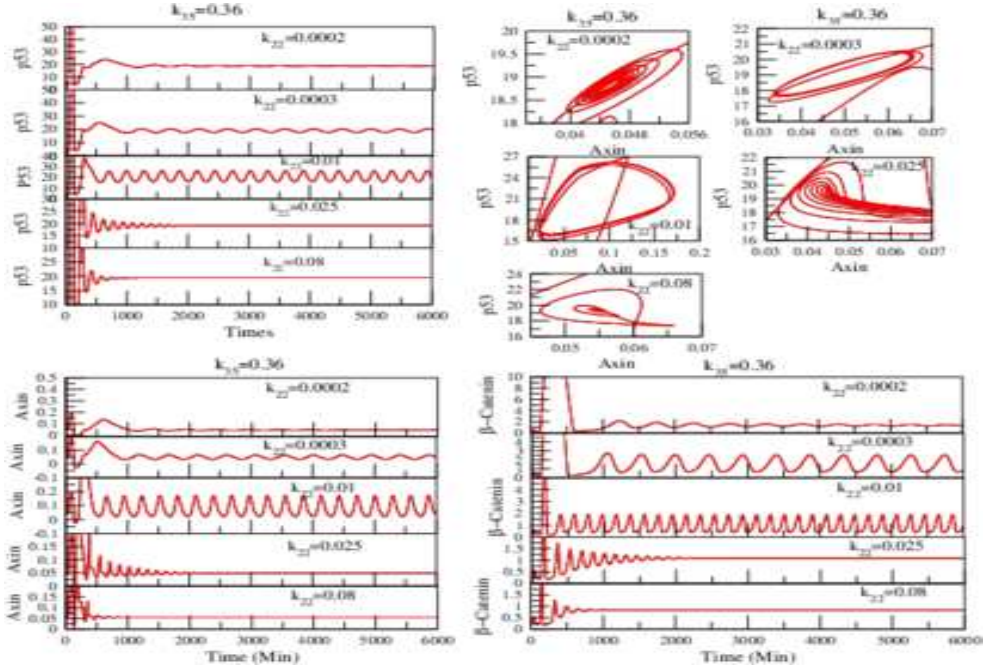}
\caption{Plots of $p53$, Axin and $\beta$-catenin as a function of time for $k_{22}$ values 0.0002, 0.0003, 0.01, 0.025 and 0.08, for fixed values of $k_{35}$. The upper right panels shows the two dimensional plots of $p53$ and $\beta$-catenin with Axin for same set of $k_{22}$ and $k_{35}$ parameter values.} 
\end{center}
\end{figure*}

\subsection*{Dynamics of $p53$ regulated by Wnt}
The activation of Axin occurs after the $Gsk$ initiation from the signal received at the membrane. The Nutilin synthesis rate ($k_{35}$) is kept constant in this case ($k_{35}=0.36$), giving a slight stress to the system. In order to understand how does Wnt oscillator regulate $p53$, we allow to change Axin synthesis rate $k_{22}$ and see the dynamical behavior of $p53$ induced by Axin, which is one of the most important molecular species in Wnt network module. The dynamics of the Axin and $p53$ for different values of $k_{22}$ (Fig. 4). For small values of $k_{22}$ ($k_{22}<0.0002$), Axin dynamics maintains lowest constant level, showing stabilized state of the Axin dynamics (Fig. 4 lower left panels). The increase in the value of $k_{22}$ allows the Axin dynamics to switch from stabilized state to damped state for certain range of time and then become stabilized. The co-existence of these two states (damped and stabilized states) takes place for a certain range on $k_{22}$ ($0.0003<k_{22}<0.008$). If we increase the value of $k_{22}$, the duration of damped oscillation also increases. The switching of damped state to sustain oscillation state takes place for certain range of $k_{22}$, $0.0087<k_{22}<0.015$. Further increase in $k_{22}$ force the sustain oscillation state to damped oscillation state again, where the time interval of damped oscillation becomes smaller as $k_{22}$ increases. If we increase $k_{22}$ value, then Axin dynamics landed to stabilized state.

We then show the dynamics of $\beta-$ Catenin as a function of $k_{22}$ (Fig. 4) which indicates the switching of the three different states driven by $k_{22}$. In this case we could able to find sustain oscillation very quickly indicating that the $\beta-$Catenin can be quickly switch to stress condition and easily. Similarly, $\beta-$Catenin can reach second stabilized state (probably cycle arrest) quickly and easily as compared to other molecular species. The dynamical behavior of $\beta-$Catenin is supported by two dimensional plots of $\beta-$Catenin and Axin (Fig. 3 middle right panels) which exhibit fixed point oscillations (corresponding to two stabilized states), attractor towards the stable fixed points (corresponding to two damped oscillations), limit cycle (corresponding to sustain oscillation). 

Now the dynamics of $p53$ as a function of $k_{22}$ (Fig. 4 upper left panel) show that the patterns in the dynamics of molecular species in Wnt oscillator are well captured in the $p53$ dynamics. The $p53$ dynamics exhibit the corresponding states, namely, stabilized, damped and sustain oscillation states, as observed in Axin and $\beta-$Catenin as a function of $k_{22}$ with similar patterns and dynamics.
\begin{figure*}
\label{fig5}
\begin{center}
\includegraphics[height=270pt,width=13.0cm]{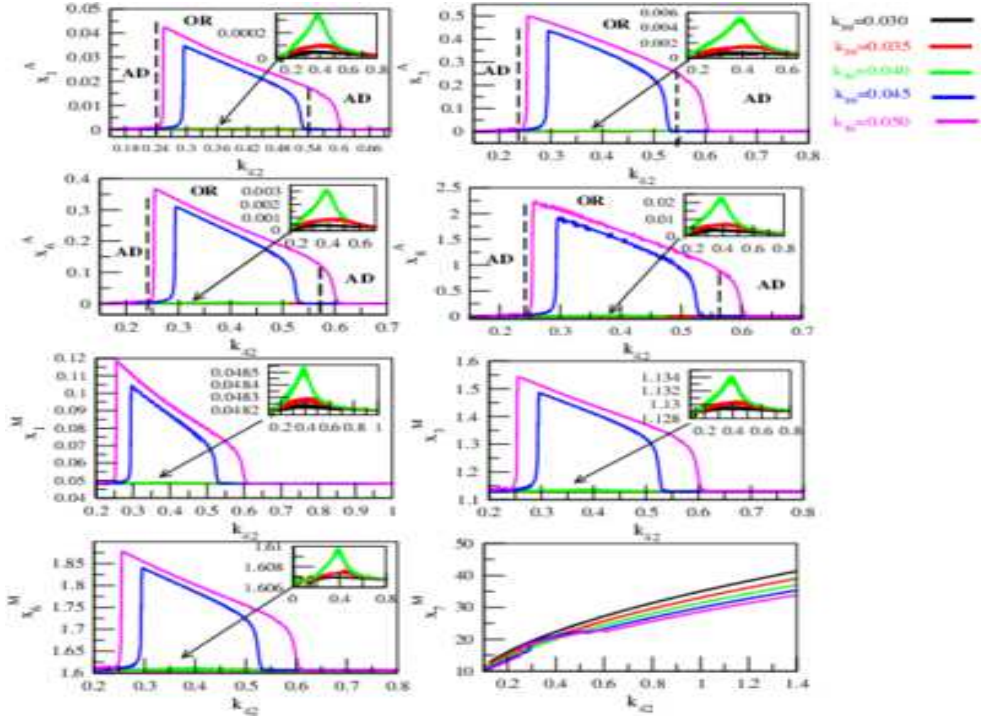}
\caption{The amplitudes of $p53~(x_1^A)$, $\beta$-catenin ($x^A_3$), Gsk3 ($x^A_6$) and $Mdm2$ ($x^A_8$) as a function of $k_{35}$ for various $k_{39}$ values 0.030, 0.035, 0.040, 0.045 and 0.050 (upper four panels). The lower four panels are the maxima of $p53~(x_1^M)$, $\beta$-catenin ($x^M_3$), Gsk3 ($x^M_6$) and $p53$ ($x^M_7$) for the set of parameter values.} 
\end{center}
\end{figure*}

The important dynamical behaviors those have been captured in $p53$ dynamics due to change of states in molecular species in Wnt oscillator are synchronous in states and time periods of dynamics. If we look at the dynamics of Axin and $\beta-$Catenin as a function of $k_{22}$, we can easily observe two important changes, (1) change in time period of oscillations of the molecular species, and (2) change in the pattern of states. The increase in $k_{22}$ force to decrease the time period of the Axin and $\beta-$Catenin dynamics. At the same time the switching of the various states takes place. These changes in the Axin and $\beta-$Catenin are systematically and synchronously well captured in qualitative sense in the $p53$ dynamics (Fig. 4) during the communication of the two $p53$ and Wnt oscillators via $Gsk3$.

\subsection*{Amplitude death driven by $Gsk3$ and $p53$ interaction}
We now study the impact of interaction of $Gsk3$ and $p53$ ($k_{46}$) on Wnt and $p53$ cross talk driven by Nutlin concentration level in the system (Fig. 5). The rate of interaction of $p53$ and $Gsk3$ can modulated by external catalytic action or external stimuli regulating the interaction in the system. 

\subsubsection*{Wnt cycle arrest induced by Nutlin}
We calculated amplitudes of four molecular species Axin, $\beta-$Catenin, $Gsk3$ and $Mdm2$ as a function of $k_{35}$ for five different values of $k_{46}$ (Fig. 5 upper two rows). The amplitudes of the molecular species  for every values of $k_{42}$ are calculated by averaging the amplitudes within the interval of time $[100,1500]$ minutes. From the results we observe three different regimes for each value of $k_{46}$: first amplitude death (AD) regime (for small $k_{35}$); second finite amplitude or oscillation regime (OR) (for moderate $k_{35}$); and third amplitude death regime again (for large $k_{35}$). The first AD regime may correspond to normal state (stable state) of the system, and second AD regime may correspond to apoptosis (cycle arrest) of the system. Further, we calculated maximum values of molecular species Axin, $\beta-$Catenin, $Gsk3$ and $p53$ for the correponding values taken above (Fig. 6 lower two rows of plots). The patterns obtained in these plots are similar to the patterns of the amplitude plots of the corresponding molecular species.
\begin{figure*}
\label{fig6}
\begin{center}
\includegraphics[height=280pt,width=13.0cm]{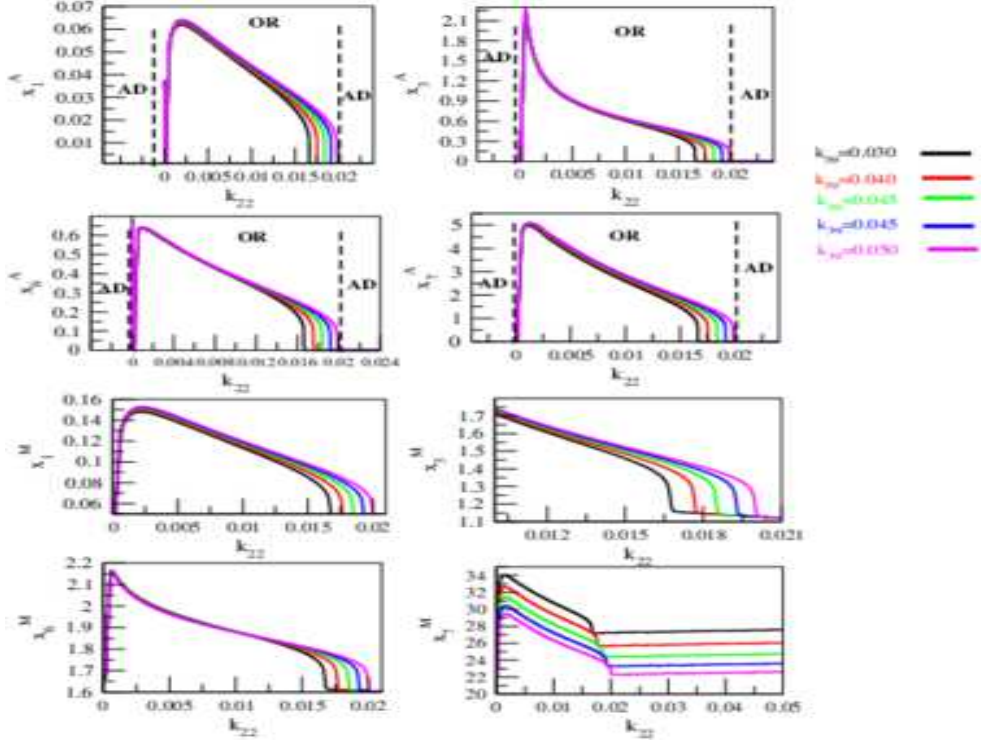}
\caption{Similar plots of $p53~(x_1)$, $\beta$-catenin ($x_3$), Gsk3 ($x_6$) and $p53$ ($x_7$) as a function of $k_{22}$ for the same $k_{39}$ values as in Figure 5.} 
\end{center}
\end{figure*}

In OR, the amplitude is finite and increases monotonically as a function of $k_{42}$, then reaches a maximum amplitude, after which amplitude decreases slowly and then decreases monotonically. This is the regime where the system is in activated or stress condition. It is also to be noted that as $k_{46}$ increases the amplitude also increases and the range of activated regime also increases. This means that increasing $k_{46}$ force the system to get activated quickly and protects the system from next stabilized regime (cycle arrest or apoptosis regime). Therefore, even though increase in Nutlin concentration drives the system to apoptosis regime quickly, one can modulate the interaction of $p53$ and $Gsk3$ to protect the system from apoptosis upto significant range of Nutlin concentration level, so that the system can repair back from toxic due to excess Nutlin to come back to normal condition.

\subsubsection*{$p53$ cycle arrest induced by Axin}
Now we study the impact of $p53$ and $Gsk3$ interaction on the condition of activation of molecular species of both the $p53$ and Wnt oscillators. We calculated the amplitudes of these molecular species as a function of Axin concentration level in the system ($k_{22}$) (Fig. 6 first two rows) for the same parameter values as we had taken in the case of calculation of amplitudes of these molecular species with the variation of Nutlin concentration in the system. For the same parameter values we also calculated the maxima of these molecular species (Fig. 6 lower two rows).

The change in $k_{22}$ drives the amplitudes of all the molecular species to different oscillation states (AD, OR, AD) as we have obtain in the case of Wnt cycle arrest by $k_{42}$ variation. The increase in Axin concentration level in the system forces amplitudes of the molecular species variables to pass the trajectories of first AD (normal condition), then OR (stress condition) and then second AD (cycle arrest condition). However, in this case the amplitudes of the variables do not change much as a function $k_{22}$, but significant change in the range of second AD as a function of $k_{46}$. The results indicate the Axin induced $p53$ cycle arrest, however, interaction of $p53$ and $Gsk3$ can able to protect the system from apoptosis upto significant range of $k_{22}$ so that the system can repair back to its normal condition.

\subsection*{Steady state solution of the system}
Since the direct and exact analytical solution of the set of coupled nonlinear differential equations (\ref{ode}), we used quasi steady state approximation \cite{Mur,Sch} to obtain approximation solution of the system. Using steady state values of the fast variables using (\ref{steady}), putting back to the ODE of $x_1(t)$ in equation (\ref{reduce}) and taking $\frac{x_1}{k_{24}+x_1}$$\sim\frac{1}{1+\frac{k_{24}}{x_1}}$$\sim\left(1-\frac{k_{24}}{x_1}\right)\sim 1$ (for large $x_1$) we get the following approximate solutions of $x_1$,
\begin{eqnarray}
\label{x1}
x_1(t)\approx\frac{u_1}{u_2}+C_1e^{-u_2t}
\end{eqnarray}
where, $C_1$ is a constant, and $u_1$, $u_2$ are also given by,
\begin{eqnarray}
\label{sx1}
u_1&=&k_{26}\left(k_{22}x_2^*+k_2k_8-\frac{u_2k_2}{k_1k_{26}}-k_{23}\right)\nonumber\\
u_2&=&\frac{k_{1}k_{26}k_{40}x_{13}^*}{\frac{k_{39}}{k_{28}}\left(k_{41}x_{13}^*-k_{29}x_9^*\right)}
\end{eqnarray}
The equation (\ref{x1}) shows that $x_1$ becomes constant (steady state) as $t\rightarrow\infty$. However, $x_1$ depends on various other steady state variables as shown in equation (\ref{x1}) and (\ref{sx1}). 

We then solve ODE of $x_3(t)$ in equation (\ref{reduce}) for $x_3$, and the result is given by,
\begin{eqnarray}
\label{x3}
x_3(t)\approx \frac{k_3-k_{14}x_4^*}{k_{13}}+C_2e^{-k_{13}k_{26}t}
\end{eqnarray}
where, $C_2$ is a constant. Here, in this solution, we can get that $x_3$ stabilizes to a constant value at sufficiently large time limit ($t\rightarrow\infty$).

Now, the approximate solution of $x_6$ can be obtained by solving ODE of $x_6$ in equation (\ref{reduce}), using equation (\ref{x1}) and (\ref{equi}), as given below,
\begin{eqnarray}
x_6(t)\approx exp\left(at-be^{-u_2t}\right)\left[C_3-\left(\frac{u_3}{u_2}\right)b^{a/u_2}\Gamma\left(-\frac{a}{u_2},b\right)\right]
\end{eqnarray}
where, $C_3$ is a constant. Similarly, $u_3=k_2k_8k_{26}$, $a=\frac{k_1k_{26}u_1}{u_2}$ and $b=\frac{k_1k_{26}C_1}{u_2}$ are also constants.

The molecular species $x_{11}$ can be solved directly from equation (\ref{reduce}) and using equation (\ref{equi}). The result is given by,
\begin{eqnarray}
\label{x11}
x_{11}(t)\approx C_4e^{-k_{38}t}+\frac{k_{35}}{k_{38}}
\end{eqnarray}
where, $C_4$ is a constant. Further, using the differential equation in $x_{7}$ in equation (\ref{reduce}) and equation (\ref{x11}) along with equation (\ref{equi}), we solve for $x_7$ given by,
\begin{eqnarray}
\label{x7}
x_{7}(t)&\approx& \frac{\frac{u_4}{k_{38}}\int\frac{y^{u_5/k_{38}t}}{y-C_4}dy+C_5}{\left[C_4+\frac{k_{35}}{k_{38}}e^{k_{38}t}\right]}\nonumber\\
&\approx& \frac{2\pi i\frac{u_4}{k_{38}}C_4^{\frac{u_{5}}{k_{38}}}+C_5}{\left[C_4+\frac{k_{35}}{k_{38}}e^{k_{38}t}\right]}
\end{eqnarray}
where, $u_4=k_{31}+k_{34}x^*_{10}$, $u_5=\frac{1}{k_{36}}(k_{33}k_{37}x^*_{12})$ and $C_5$ is a constant. Similarly, the variable $x_8$ can be solved using equations (\ref{reduce}), (\ref{x11}) and (\ref{equi}) as follows,
\begin{eqnarray}
\label{x8}
x_8(t)&\approx&\frac{C_6-\frac{k_{27}x^*_9}{k_{38}}\int y^{-\left(1+\frac{u_5}{k_{38}}\right)e^{-\frac{u_7}{k_{38}}y}}dy}{e^{-u_6t+\frac{u_7}{k_{38}}e^{-k_{38}t}}}\nonumber\\
&\approx&\frac{C_6-\frac{k_{27}x^*_9}{k_{38}}E_{1+\frac{u_7}{k_{38}}}\left(\frac{u_7}{k_{38}}\right)}{e^{-u_6t+\frac{u_7}{k_{38}}e^{-k_{38}t}}}
\end{eqnarray}
where, $u_6=k_{38}+\frac{k_{35}k_{36}}{k_{38}}$, $u_7$=$k_{36}C_1$ and $C_6$ are constants. $E$ is the error function. 

\section*{Conclusion}
The cross talk between $p53$ and Wnt oscillators can influence each other such that the dynamics one oscillator can regulate the dynamics of the other and vice versa. The concentration level of Nutlin can drive the $p53$ network module to different oscillating states corresponding to different stress states of the system. This changes in the oscillating states can be well communicated to Wnt network module via $Gsk3$ and could able to regulate the Wnt functionalities. Further, this changes can well interfere and give impact on the process of sometogenesis of Wnt network module. On the other hand, the available concentration level of Axin in Wnt network module can also influence the dynamics of molecular species variables in Wnt network module at different stress states. Then the information of these changes is well process by the $p53$ network module via $Gsk3$ and capture the changes accordingly. This means that the $p53$ and Wnt network modules can cross talk between them and regulate each other due to changes in each oscillator by processing information between them. However, excess changes in each network module can act as toxic to each module and arrest the cycle of the other, and vice versa.

We then demonstrate that any other changes in the regulation of the network module, may probably by catalytic reaction or external stimuli which alters the rate of the reaction of a particular reaction, can probably try to protect the system from apoptosis upto certain range of system's condition within which the system can repair back the changes to come back to normal condition. We attempt to numerically demonstrate this phenomena by allowing the change in the rate constant of $p53$ and $Gsk3$ interaction and clearly give a chance to both the $p53$ and Wnt network modules upto certain range of excess concentration levels of Nutlin and Axin to protect themselves from cycle arrest, namely apoptosis. This means that one important role of catalytic reaction and external stimuli (which do not affect the reactions and network but rate of reactions) is to protect the system from apoptosis or system's failure by giving a chance of repair the failure or defect by itself within the system automatically. Therefore, we need to find out various such reactions and stimuli in cellular network in order to control and prevent the cell from cell death, due to cellular breakdown by diseases or other external mechanisms, by regulating these important reactions and stimuli.

\pagebreak
\begin{table*}
\begin{center}
{\bf Table 1-List of molecular species and their intial concentration} 
\begin{tabular}{|l|p{3cm}|p{6.5cm}|p{1.5cm}|p{2.5cm}|}
 \hline \multicolumn{5}{}{} \\ \hline

\bf{ S.No.}& \bf{Species Name}	&   \bf{Description} 	& \bf{Notation}	&\bf{Intial Concentration(nM)}	\\ \hline
1.         & $Axin2$        	& Intial Concentration $Axin2$ protein 	& $x_1$ 	&$10.0$		\\ \hline
2.         & $Axin2$\_mRNA      & Intial Concentration $Axin2$\_mRNA	& $x_2$ 	&$0.1$		\\ \hline
3.         & $\beta$-catenin    & UnPhosphorylated $\beta$-catenin	& $x_3$ 	&$0.1$		\\ \hline
4.         & $\beta$-catenin\_p & Phosphorylated $\beta$-catenin	& $x_4$  	&$0.001$	\\ \hline
5.         & $\beta$-catenin\_n & Nuclear $\beta$-catenin   		& $x_5$  	&$0.1$		\\ \hline
6.         & $Gsk3$             & Gsk3 protien       			& $x_6$  	&$5.0$		\\ \hline
7.         & $p53$	        & unbound $p53$ protein  		& $x_7$  	&$0.022$	\\ \hline
8.         & $MDM2$             & Unbound $MDM2$ protien       		& $x_8$  	&$0.035$	\\ \hline
9.         & $MDM2$\_mRNA       & $MDM2$ Messsenger mRNA        	& $x_9$  	&$0.01$		\\ \hline
10.        & $p53$\_$MDM2$      & $MDM2$ and $p53$ complex      	& $x_{10}$  	&$0.01$		\\ \hline
11.        & Nutlin             & Unbound Nutlin			&$x_{11}$  	&$0.0425$	\\ \hline
12.        & Nutlin\_$MDM2$	& Nutlin $MDM2$ complex      		& $x_{12}$  	&$0.01$		\\ \hline
13.        & $p53$\_$Gsk3$	& $p53$ and $Gsk3$ complex        	& $x_{13}$  	&$0.01$		\\ \hline
\end{tabular}
\end{center}
\end{table*}

\begin{table*}
\begin{center}
{\bf Table 2 List of parameter} 
\begin{tabular}{|l|p{1.5cm}|p{7.5cm}|p{2cm}|p{1.5cm}|}
 \hline \multicolumn{5}{}{} \\ \hline

\bf{ S.No.}& \bf{Notation}& \bf{Description} 							& \bf{Values}		& \bf{Reference}\\ \hline
1. & $K_1$ & Rate constant for binding of $Gsk3$ to $Axin2$  					&$0.23nMmin^{-1}$  	&\cite{Gol}	\\ \hline
2. & $K_2$ & Rate constant for dissociation of $Gsk3$\_$Axin2$  				&$0.1min^{-1}$ 		&\cite{Gol}	\\ \hline
3. & $K_3$ & Rate of $\beta$-catenin synthesis 		    					&$0.087nMmin^{-1}$	&\cite{Gol}	\\ \hline
4. & $K_4$ & Rate of $\beta$-catenin entry into the nucleus					&$0.7min^{-1}$  	&\cite{Gol}	\\ \hline
5. & $K_5$ & Rate of $\beta$-catenin exit from the nucleus					&$1.5min^{-1}$  	&\cite{Gol} 	\\ \hline
6. & $K_6$ & Rate of phosphorylation of $\beta$-catenin by the $Gsk3$				&$5.08nMmin^{-1}$	&\cite{Gol}	\\ \hline
7. & $K_7$ & Concentration of Dishevelled(Dsh)protein		  				&$2.0nM$	        &\cite{Gol} 	\\ \hline
8. & $K_8$ & Total $Gsk3$ Concentration				       				&$3.0nM$	        &\cite{Gol}	\\ \hline
9. & $K_9$ & Rate of inhibition by Dsh for$\beta$-catenin\_p by the $Axin2$\_$Gsk3$ complex	&$0.5nM$ 		&\cite{Gol} 	\\ \hline
10.& $K_{10}$ & Rate constant for $\beta$-catenin\_p by the $Axin2$\_$Gsk3$ complex		&$0.28nM$  		&\cite{Gol} 	\\ \hline
11.& $K_{11}$ &Rate constant of dephosphorylation of $\beta$-catenin				&$1.0nMmin{^-1}$ 	&\cite{Gol}	\\ \hline
12.& $K_{12}$& Maximun rate constant for $\beta$-catenin phosphoration 		       		&$0.003nM$  		&\cite{Gol}	\\ \hline
13.& $K_{13}$& Rate constant for degradation of unphosphorylated $\beta$-catenin  		&$0$  	   		&\cite{Gol}	\\ \hline
14.& $K_{14}$&Rate constant for degradation of phosphorylated $\beta$-catenin	  		&$7.062min{-1}$  	&\cite{Gol} 	\\ \hline
15.& $K_{15}$&Rate of transcription of the $Axin2$ gene				  		&$0.06nMmin^{-1}$	&\cite{Gol} 	\\ \hline
16.& $K_{16}$&Rate of transcription of the $Axin2$ gene induced by nuclear $\beta$-catenin	&$1.64nMmin^{-1}$	&\cite{Gol}	 \\ \hline
17.& $K_{17}$&Rate for induction by nuclear $\beta$-catenin of $Axin2$ gene trascription 	&$0.7nM$    		&\cite{Gol} 	\\\hline
18.& $K_{18}$&Maximum Rate of degradation of $Axin2$\_mRNA					&$0.8nMmin^{-1}$ 	&\cite{Gol} 	\\ \hline
19.& $K_{19}$&Rate constant for degradation of $Axin2$\_mRNA					&$0.48nM$		&\cite{Gol} 	\\ \hline 
20.& $K_{20}$&Rate of transcrioption of $Axin2$ gene induced by transcription factor  		&$0.5nMmin^{-1}$ 	&\cite{Gol} 	\\ \hline 
21.& $K_{21}$&rate for induction by transcription factor of $Axin2$ gene transcription  	&$0.05nM$ 		&\cite{Gol} 	\\ \hline 
22.& $K_{22}$&Rate of synthesis of $Axin2$ protein						&$0.02min^{-1}$  	&\cite{Gol} 	\\ \hline 
23.& $K_{23}$&Maximum rate of degradtion of $Axin2$ protein					&$0.6nMmin^{-1}$ 	&\cite{Gol} 	\\ \hline 
24.& $K_{24}$& Michaelis rate for degradtion of $Axin2$ protein 		     		&$0.63nM$		&\cite{Gol}	\\ \hline 
25.& $K_{25}$&Hill coefficients									&$2.0$			&\cite{Gol}	\\ \hline 
26.& $k_{26}$&Scaling factor for Wnt oscillator							&$1.5$			&\cite{Gol} 	\\ \hline
27.&$k_{27}$&$MDM2$ synthesis									&$0.297min^{-1}$ 	&\cite{Pro} 	\\ \hline
28.&$k_{28}$&$MDM2$ transcripton								&$0.006min^{-1}$ 	&\cite{Pro}	\\ \hline	
29.&$k_{29}$&$MDM2_mRNA$ degradation								&$0.006min^{-1}$ 	&\cite{Pro}	\\ \hline	
30.&$k_{30}$&$MDM2$ degradation									&$0.2598min^{-1}$ 	&\cite{Pro}	\\ \hline
31.&$k_{31}$&$p53$ synthesis									&$4.68min^{-1}$		&\cite{Pro}	\\ \hline
32.&$k_{32}$&$p53$ degradation									&$0.0495min^{-1}$	&\cite{Pro}	\\ \hline
33.&$k_{33}$&$p53\_MDM2$ binding								&$0.693min^{-1}$ 	&\cite{Pro}	\\ \hline
\end{tabular}
\end{center}
\end{table*}

\begin{table*}
\begin{center}
{\bf Table 2 Continue...} 
\begin{tabular}{|l|p{1.5cm}|p{7.5cm}|p{2cm}|p{1.5cm}|}
 \hline \multicolumn{5}{}{} \\ \hline

\bf{ S.No.}& \bf{Notation}& \bf{Description} 			& \bf{Values}		& \bf{Reference}	\\ \hline
34.&$k_{34}$&$p53\_MDM2$ dissociation				&$0.00693min^{-1}$	&\cite{Pro}		\\ \hline
35.&$k_{35}$&Nutilin formation					&$0.001min^{-1}$	&Estimated		\\ \hline
36.&$k_{36}$&Nutilin $MDM2$ formation                      	&$0.012min^{-1}$	&Estimated		\\ \hline		
37.&$k_{37}$&Nutilin $MDM2$ degradation				&$0.03min^{-1}$		&Estimated		\\ \hline
38.&$k_{38}$&Nutilin degaradation				&$0.06min^{-1}$		&Estimated		\\ \hline
39.&$k_{39}$&$p53\_Gsk3$ complex formation			&$0.04min^{-1}$		&\cite{Proc}		\\ \hline
40.&$k_{40}$&$p53\_Gsk3$ complex degradation			&$0.12min^{-1}$		&\cite{Proc}		\\ \hline
41.&$k_{41}$&$MDM2\_mRNA$ synthesis				&$0.042min^{-1}$	&\cite{Proc}		\\ \hline
\end{tabular}
\end{center}
\end{table*}

\pagebreak

\end{document}